\documentclass[aps,twocolumn,prc,superscriptaddress,showpacs,nofootinbib,floatfix,amssymb,amsfonts,amsmath]{revtex4-1}
\usepackage{graphicx}
\usepackage{epsf}
\usepackage{amsmath,amssymb}
\usepackage{bm}
\usepackage{color}
\usepackage{ulem}

\newcommand{\nuc}[2]{{}^{#2} \mathrm{#1}}
\begin{document}
\title{
  Consistent description for cluster dynamics and single-particle correlation
}%
\author{Naoyuki Itagaki}
\affiliation{
  Yukawa Institute for Theoretical Physics, Kyoto University,
  Kitashirakawa Oiwake-Cho, Kyoto 606-8502, Japan
}
\author{Tomoya Naito}
\affiliation{
  Department of Physics, Graduate School of Science, The University of Tokyo, Tokyo 113-0033, Japan}
\affiliation{
  RIKEN Nishina Center, Wako 351-0198, Wako 351-0198, Japan
}
\date{\today}
\preprint{RIKEN-QHP-485}
\preprint{RIKEN-iTHEMS-Report-20}
\begin{abstract}
  Cluster dynamics and single-particle correlation are simultaneously treated for the description of the ground state of $\nuc{C}{12}$.
  The recent development of the antisymmetrized quasi cluster model (AQCM) makes it possible to
  generate $jj$-coupling shell-model wave functions 
  from $\alpha$ clusters models.
  The cluster dynamics and the competition with the 
  $jj$-coupling shell-model structure can be estimated rather easily.
  In the present study, we further include the effect of single-particle excitation;
  the mixing of the two-particle-two-hole excited states is considered.
  The single-particle excitation is not always taken into account in the standard cluster model analyses,
  and the two-particle-two-hole states are found to strongly contribute to the lowering 
  of the ground state owing to the pairing-like correlations.
  By extending AQCM, all of the basis states are prepared on the same footing, and they are superposed based on
  the framework of the generator coordinate method (GCM).
\end{abstract}
\maketitle
%
\section{Introduction}
\par
The $\nuc{He}{4}$ nuclei have been known to have large binding energy in the light mass region.
On the contrary, the relative interaction between $\nuc{He}{4}$ nuclei is weak.
Therefore, they can be subsystems called $\alpha$ clusters
in some of light nuclei~\cite{Brink,RevModPhys.90.035004}. 
The search for the candidates for  the $\alpha$ cluster structure has been performed for decades,
and the most famous example is the second $0^+$ state of $\nuc{C}{12}$
called Hoyle state, which has a developed three-$\alpha$ cluster structure~\cite{Hoyle1954,FREER20141}.
The cluster models have been found to be capable of describing various properties of the Hoyle state~\cite{PTPS.68.29,PhysRevLett.87.192501}.
\par
In most of the conventional cluster models, the clusters treated as subsystems
have been limited to nuclei corresponding to the closure
of the three-dimensional harmonic oscillator, such as $\nuc{He}{4}$, $\nuc{O}{16}$ and $\nuc{Ca}{40}$.
In these cases, the contribution of the non-central interactions (spin-orbit and tensor interactions) vanishes.
This is because
the closure configurations of the major shells can create only spin zero systems owing to the antisymmetrization effect. 
This point was the big problem of these traditional cluster models;
the non-central interactions
work neither inside clusters nor between $\alpha$ clusters.
The spin-orbit  interaction
is known to be quite important in nuclear systems, especially 
in explaining the observed magic numbers;
the subclosure configurations of 
the $jj$-coupling shell model
($f_{7/2}$, $g_{9/2}$, and $h_{11/2}$)
correspond to the observed magic numbers of $28$, $50$, and $126$~\cite{Mayer}.
Indeed this spin-orbit interaction is 
known to work as a driving force to break the
traditional clusters
corresponding to the closures of the major shells,
when the model space is
extended and the path to another symmetry is opened~\cite{PhysRevC.70.054307}.
\par
To include the spin-orbit contribution
in the theoretical model starting with the traditional cluster model side,
we proposed the antisymmetrized quasi cluster model
(AQCM)~\cite{PhysRevC.94.064324,PhysRevC.73.034310,PhysRevC.75.054309,PhysRevC.79.034308,PhysRevC.83.014302,PhysRevC.87.054334,ptep093D01,ptep063D01,ptepptx161,PhysRevC.97.014307,PhysRevC.98.044306,PhysRevC.101.034304,PhysRevC.102.024332}.
This method allows us to smoothly transform $\alpha$ cluster model wave functions to
$jj$-coupling shell model ones, and
we call the clusters that feel the effect of  the spin-orbit interaction owing to this model quasi clusters.
In AQCM,
we have two parameters: $R$ representing the distance between $\alpha$ clusters
and $\Lambda$ characterizing the transition of $\alpha$ cluster(s) to quasi cluster(s).
The $jj$-coupling shell model states can be obtained starting with the $\alpha$ cluster model
by changing $\alpha$ clusters to quasi clusters (giving finite $\Lambda$ values to $\alpha$ clusters)
and taking small distance limit of $R$. 
It has been known that the conventional $\alpha$ cluster models cover the model space of closure of major shells
($N=2$, $N=8$, $N=20$, \textit{etc.}).
In addition, by changing $\alpha$ clusters to quasi clusters,
the subclosure configurations of the $jj$-coupling shell model,
$p_{3/2}$ ($N=6$), $d_{5/2}$ ($N=14$), $f_{7/2}$ ($N=28$), and $g_{9/2}$ ($N=50$),
which arise from the spin-orbit interaction in the mean-field,
can be described by our AQCM~\cite{ptep093D01}.
\par
We have previously introduced AQCM to $\nuc{C}{12}$ and discussed the
competition between the cluster states and $jj$-coupling shell model state~\cite{PhysRevC.94.064324}.
The consistent description of $\nuc{C}{12}$ and $\nuc{O}{16}$, which has been a long-standing problem
of microscopic cluster models, has been achieved.
In this paper, we examine again $\nuc{C}{12}$, 
where not only the competition between the
cluster states and the lowest shell-model configuration,
the effect of single-particle excitation is further included 
for the description of the ground state.
The mixing of the two-particle-two-hole excited states 
owing to the pairing-like correlations is examined.
By extending AQCM, all of the basis states are prepared on the same footing, and they are superposed based on
the framework of the generator coordinate method (GCM).
Although it has been analytically shown to be feasible to prepare some of the two-particle-two-hole configurations of
the $jj$-coupling shell model within the framework of AQCM~\cite{ptepptx161},
here we try to use much simpler method.
The two-particle-two-hole states 
around the optimal AQCM basis state are generated using numerical technique.
Owing to the generation of many states compared with our previous approach,
much larger effect for the lowering of the energy
due to the mixing of two-particle-two-hole states 
will be discussed.
\par
The nucleus of $\nuc{C}{12}$ is the typical example which has both characters of cluster and shell aspects.
Recently, various kinds of microscopic approaches have shown the importance of
the mixing of shell and cluster components. Not only the energy levels,
various properties including electromagnetic transition strengths,
$\alpha$-decay widths, and scattering phenomena have been discussed~\cite{
  PTP.117.655,
  PhysRevLett.106.192501,
  PhysRevLett.109.252501,
  PhysRevC.91.024315,
  PhysRevC.93.054307,
  PhysRevLett.98.032501,
  PhysRevLett.105.022501,
  PhysRevC.36.54,
  DESCOUVEMONT2002275,
  DREYFUSS2013511,
  PhysRevLett.113.012502,
  PhysRevLett.111.092501,
  PhysRevLett.118.152503,
  Launey2018}.
Especially, based on the antisymmetrized molecular dynamics (AMD), the one-particle-one-hole states
are discussed in relation with the isoscalar monopole and dipole resonance strengths~\cite{PhysRevC.93.054307}.
In this approach, one-particle-one-hole states are expressed by the small shift of one particle
around the optimal AMD solution. 
Here in our study, we focus on the two-particle-two-hole excitation,
which covers the model space of one-particle-one-hole excitation, and 
the lowering of the energy owing to  the effect of BCS-like paring can be clarified.
Some of the preceding works are based on  modern \textit{ab initio} approaches, where the tensor and
short-range correlations are included. 
Compared with these, our approach is rather phenomenological, but here
we examine the natural extension of the 
AQCM framework and include both cluster dynamics and the single-particle excitation. 
\par
This paper is organized as follows. 
The framework is described in  Sec.~\ref{Frame}.
The results are shown in Sec.~\ref{Results}.
The conclusions are presented in Sec.~\ref{Concl}.
%
\section{framework}
\label{Frame}
%
\subsection{Basic feature of AQCM}
\par
AQCM 
allows the smooth transformation of the cluster model 
wave functions to the $jj$-coupling shell model ones.
In AQCM, each single particle is described by a Gaussian form
as in many other cluster models including the Brink model~\cite{Brink},
\begin{equation}	
  \phi^{\tau, \sigma} \left( \bm{r} \right)
  =
  \left(  \frac{2\nu}{\pi} \right)^{\frac{3}{4}} 
  \exp \left[- \nu \left(\bm{r} - \bm{\zeta} \right)^{2} \right] \chi^{\tau,\sigma}, 
  \label{spwf} 
\end{equation}
where the Gaussian center parameter $\bm{\zeta}$
is related to the expectation 
value of the position of the nucleon,
and $\chi^{\tau,\sigma}$ is the spin-isospin part of the wave function.
For the size parameter $\nu$, 
here we use $\nu = 0.23 \, \mathrm{fm}^{-2}$, which gives the optimal $0^+$ energy
of $\nuc{C}{12}$ within a single AQCM basis state.
The Slater determinant is constructed from 
these single-particle wave functions by antisymmetrizing them.
\par
Next we focus on the Gaussian center parameters
$\left\{ \bm{\zeta}_i \right\}$.
As in other cluster models, here four single-particle 
wave functions with different spin and isospin
sharing a common 
$\bm{\zeta}$ value correspond to an $\alpha$ cluster.
This cluster wave function is transformed into
$jj$-coupling shell model based on the AQCM.
When the original value of the Gaussian center parameter $\bm{\zeta}$
is $\bm{R}$,
which is 
real and
related to the spatial position of this nucleon, 
it is transformed 
by adding the imaginary part as
\begin{equation}
  \bm{\zeta} = \bm{R} + i \Lambda \bm{e}^{\text{spin}} \times \bm{R}, 
  \label{AQCM}
\end{equation}
where $\bm{e}^{\text{spin}}$ is a unit vector for the intrinsic-spin orientation of this
nucleon. 
The control parameter $\Lambda$ is associated with the breaking of the cluster,
and with a finite value of $\Lambda$, the two nucleons with opposite spin orientations 
have the $\bm{\zeta}$ values, which are complex conjugate  with each other.
This situation corresponds to the time-reversal motion of two nucleons.
After this transformation, the $\alpha$ clusters are called quasi clusters.
\par
Here we explain the intuitive meaning of this procedure.
The inclusion of the imaginary part allows us to directly connect 
the single-particle wave function
to the spherical harmonics of the 
$jj$-coupling shell model.  
Suppose that the Gaussian center parameter $\bm{\zeta}$ has the $x$ component, 
and the spin direction is defined along the $z$ axis (this is spin-up nucleon). 
According to Eq.~(\ref{AQCM}), 
the imaginary part of $\bm{\zeta}$ is given to its $y$ component. 
When we expand $-\nu \left(\bm{r} - \bm{\zeta} \right)^{2}$
in the exponent of Eq.~(\ref{spwf}), 
a factor 
$\exp \left[ 2 \nu \bm{\zeta} \cdot \bm{r} \right]$
corresponding to the cross term of this expansion
appears.
The factor 
$\exp \left[ 2 \nu \bm{\zeta} \cdot \bm{r} \right]$
contains all the information of the angular momentum of this single particle.
The Taylor expansion allows us to show that
the $p$ wave component of $\exp \left[ 2 \nu \bm{\zeta} \cdot \bm{r} \right]$ 
is $2 \nu \bm{\zeta} \cdot \bm{r}$,
which is proportional to $\left( x + i \Lambda y \right)$. 
At $\Lambda = 1$, this is proportional to $Y_{11}$ of the spherical harmonics.
The nucleon is spin-up, and thus
the coupling with the spin part gives the stretched state of the angular momentum,
$ \left| 3/2\ 3/2 \right\rangle $ of the $jj$-coupling shell model, 
where the spin-orbit interaction acts attractively. 
For the spin-down nucleon, we introduce the complex conjugate $\bm{\zeta}$ value,
which gives $ \left| 3/2\ -3/2 \right\rangle $. 
\par
In the case of $\nuc{C}{12}$, we prepare three quasi clusters.
The next two nucleons are generated by rotating the $\bm{\zeta}$ values and spin-directions of these two nucleons by $2\pi/3$. 
The last two nucleons are generated by changing the rotation angle to $4\pi/3$. 
Eventually, all the six nucleons have spin-stretched states, 
and after the antisymmetrization, the configuration becomes the subclosure configuration of 
$ \left( s1/2 \right)^2 \left( p3/2 \right)^4$.
This procedure is applied for both proton and neutron parts.
The detail is shown in Ref.~\cite{PhysRevC.87.054334}. 
\subsection{Standard AQCM for $\nuc{C}{12}$ \label{subsec-AQCM}}
\par
AQCM has been already applied to $\nuc{C}{12}$ and the essential part is recaptured here.
It has been well studied that the ground state is described by three quasi clusters
with equilateral triangular symmetry.
The parameter $R$ represents the distance between $\alpha$ clusters
with an equilateral triangular configuration, thus the distance from the origin for
each $\alpha$ cluster is $R/\sqrt{3}$. 
Following Eq.~(\ref{AQCM}), 
the Gaussian center parameters of the first quasi cluster are given as
\begin{equation}
  \bm{\zeta}^{p\uparrow, n\uparrow}_1
  =
  R \left( \bm{e_x} + i \Lambda  \bm{e_y} \right)/\sqrt{3}, 
\end{equation}
for spin-up proton ($\bm{\zeta}^{p\uparrow}_1$) and neutron ($\bm{\zeta}^{n\uparrow}_1$) and
\begin{equation}
  \bm{\zeta}^{p\downarrow, n\downarrow}_1
  =
  R \left( \bm{e_x} - i \Lambda  \bm{e_y} \right)/\sqrt{3}, 
\end{equation}
for spin-down proton ($\bm{\zeta}^{p\downarrow}_1$) and neutron.($\bm{\zeta}^{n\downarrow}_1$).
Here $\bm{e_x}$ and $\bm{e_y}$ are unit vectors of the $x$- and $y$-axis, respectively. 
The spin-isospin part of the wave function are denoted as
$\chi^{p\uparrow}_1$, $\chi^{n\uparrow}_1$, $\chi^{p\downarrow}_1$, and $\chi^{n\downarrow}_1$,
for  spin-up proton,  spin-up neutron, spin-down proton, and spin-down neutron in the first quasi cluster. 
For the second and third quasi clusters, we introduce a rotation operator around the $y$ axis
$\hat{R}_y\left(\Omega\right)$.
The Gaussian center parameters of the four nucleons in the second quasi cluster
are generated by rotating the those in the first quasi cluster around the $y$ axis
by $2\pi /3$ radian;
\begin{equation}
  \bm{\zeta}^{p\uparrow, n\uparrow, p\downarrow, n\downarrow}_2
  = 
  \hat{R}_y \left(2\pi/3\right)
  \bm{\zeta}^{p\uparrow, n\uparrow, p\downarrow, n\downarrow}_1.
\end{equation}
It is important to note that the spin-isospin part 
($\chi^{p\uparrow}_2$, $\chi^{n\uparrow}_2$, $\chi^{p\downarrow}_2$, and $\chi^{n\downarrow}_2$)
also needs to be rotated as
\begin{equation}
  \chi^{p\uparrow, n\uparrow, p\downarrow, n\downarrow}_2
  = 
  \hat{R}_y \left(2\pi/3\right)
  \chi^{p\uparrow, n\uparrow, p\downarrow, n\downarrow}_1,
\end{equation}
where the axis of the spin orientation is also tilted around the $y$ axis by $2\pi/3$ radian
(but the isospin parts do not change).
The third quasi cluster is introduced by changing the rotation angle around the $y$ axis to $4\pi/3$ radian,
\begin{equation}
  \bm{\zeta}^{p\uparrow, n\uparrow, p\downarrow, n\downarrow}_3
  = 
  \hat{R}_y \left(4\pi/3\right)
  \bm{\zeta}^{p\uparrow, n\uparrow, p\downarrow, n\downarrow}_1,
\end{equation}
for the Gaussian center parameters
($\bm{\zeta}^{p\uparrow}_3$, $\bm{\zeta}^{n\uparrow}_3$, $\bm{\zeta}^{p\downarrow}_3$, and $\bm{\zeta}^{n\downarrow}_3$)
and
\begin{equation}
  \chi^{p\uparrow, n\uparrow, p\downarrow, n\downarrow}_3
  = 
  \hat{R}_y \left(4\pi/3\right)
  \chi^{p\uparrow, n\uparrow, p\downarrow, n\downarrow}_1,
\end{equation}
for the spin-isospin part ($\chi^{p\uparrow}_3$, $\chi^{n\uparrow}_3$, $\chi^{p\downarrow}_3$, and $\chi^{n\downarrow}_3$).
\par
For the values of $R$ and $\Lambda$, we introduce
$R = 0.5$, $1.0$, $1.5$, $2.0$, $2.5$, $3.0 \, \mathrm{fm}$ and
$\Lambda = 0.0$, $0.2$, $0.4$. These 18 basis states are superposed based on GCM.
\subsection{Two-particle-two-hole states of $\nuc{C}{12}$ \label{subsec-2p2h}}
\par
The innovation of the present study is the inclusion of many two-particle-two-hole states,
by which pairing-like correlation can be taken into account.
These two-particle-two-hole basis states are generated
from the optimal AQCM basis state.
It will be shown that the AQCM basis state with $R = 2.1\,\mathrm{fm}$ and $\Lambda = 0.2$
gives the lowest energy, and Gaussian center parameters of two nucleons 
in the first quasi cluster
are shifted from this basis state using the random numbers.
We consider two sets of the basis states;
shifted two particles 
in the first quasi cluster
are either protons (with spin-up and spin-down) 
or neutrons (with spin-up and spin-down).
Both of these two corresponds to the 
isovector pairing-like excitation of protons and neutrons. 
In principle, it is possible to consider the isoscalar pairing 
of proton-neutron excitation, but this effect will be shown to be small,
maybe because the proton-neutron correlation is already
included in the quasi cluster model.
Here, the distance of the shifts are giving using a random numbers $\left\{ r_i \right\}$,
which has the probability distribution $P\left(\left|r_i\right|\right)$
proportional to  $\exp \left[-\left|r_i\right| / \sigma \right]$,
\begin{equation}
  P \left( \left| r_i \right| \right) \propto \exp \left[- \left| r_i \right| / \sigma \right].
\end{equation}
The value of $\sigma$ is chosen to be $1\,\mathrm{fm} $.
After generating $\left\{ r_i \right\}$, we multiply the sign factor to each $r_i$,
which allows $r_i$ to be positive and negative with equal probability.
The shifts of all three ($x$, $y$, $z$) directions 
for the two nucleons originally in the first quasi cluster are given using random numbers generated in this way.
Importantly, the random numbers used for the proton-proton excitation
are identical to those of neutron-neutron excitation.
Therefore, the model space still keeps the room to be isoscalar,
which is achieved when the amplitude for the wave functions for
the proton excitation and neutron excitation are obtained to be identical. 
The mixing of the isovector component in the wave function
is not the numerical artefact but due to the presence of the Coulomb interaction.
\par
It is known that proton-neutron pairing is quite important in $N=Z$ nuclei~\cite{SATULA19971,
  PhysRevLett.78.3266,
  PhysRevC.87.034310,
  Sagawa_2016},
which can be probed in the same  manner.
For the basis states corresponding to the proton-neutron pairing,
the Gaussian center parameters of spin-up proton and spin-up neutron 
in one quasi cluster are randomly generated.
However, the number of basis states must to be reduced 
due to the computational time.
\subsection{Superposition of the basis states}
\par
The 18 AQCM basis states
introduced in \ref{subsec-AQCM}
($R = 0.5$, $1.0$, $1.5$, $2.0$, $2.5$, $3.0 \, \mathrm{fm}$ and
$\Lambda = 0.0$, $0.2$, $0.4$) and
$100$ two-particle-two-hole states  
introduced in \ref{subsec-2p2h}
($50$ are for proton-proton excitation and $50$ are for neutron-neutron excitation)
are superposed based on GCM.
These $118$ basis states are abbreviated to $\left\{ \Phi_i \right\}$ ($i=1$--$118$).
They are projected to the eigen states of parity and angular momentum by
using the projection operator $P_{J^\pi}^K$,
\begin{equation}
  P_{J^\pi}^K
  =
  P^\pi \frac{2J+1}{8\pi^2}
  \int d\Omega \, {D_{MK}^J}^* R \left(\Omega \right).
\end{equation}
Here ${D_{MK}^J}$ is the Wigner $D$-function 
and $R\left(\Omega\right)$ is the rotation operator
for the spatial and spin parts of the wave function.
This integration over the Euler angle $\Omega$ is numerically performed.
The operator $P^\pi$ is for the parity projection ($P^\pi = \left(1+P^r\right) / \sqrt{2}$ for
the positive-parity states, where $P^r$ is the parity-inversion operator), 
which is also performed numerically.
This angular momentum projection enables to generate different $K$ number states
as independent basis states
from each Slater determinant.
Therefore, the total wave function $\Psi_{J^\pi}$ after the $K$-mixing is denoted as
\begin{equation}
  \Psi_{J^\pi} = \sum_{i,K} c^K_i P_{J^\pi}^K \Phi_i.
\end{equation}
The coefficients $\left\{ c^K_i \right\}$ are obtained together with the energy eigenvalue $E$
when we diagonalize the norm and Hamiltonian ($H$) matrices, namely  
by solving the Hill-Wheeler equation.
Even if the number of the basis states is 118 for the $0^+$ state,
which has only $K=0$,
the dimension of the matrices for the
other $J^\pi$ states increases through the $K$-mixing process.
\subsection{Hamiltonian}
\par
The Hamiltonian consists of the kinetic energy and 
potential energy terms.
For the potential part, the interaction consists of the central 
($\hat{V}_{\text{central}}$), 
spin-orbit
($\hat{V}_{\text{spin-orbit}}$), 
and 
Coulomb terms. For the central part, the Tohsaki interaction~\cite{PhysRevC.49.1814}  
is adopted. This interaction has finite range three-body terms in 
addition to two-body terms, which is designed to reproduce both saturation 
properties and scattering phase shifts of two $\alpha$ clusters. For the spin-orbit part, 
we use the spin-orbit term of the G3RS interaction~\cite{PTP.39.91}, which is a realistic 
interaction originally  developed to reproduce the nucleon-nucleon scattering phase 
shifts.
\par
The Tohsaki interaction 
consists of two-body ($V^{\text{(2)}}$)  and three-body ($V^{\text{(3)}}$) terms:
\begin{equation}
  \hat{V}_{\text{central}}
  =
  \frac{1}{2} \sum_{i \neq j} V^{\text{(2)}}_{ij} 
  +
  \frac{1}{6} \sum_{i \neq j, j \neq k, i \neq k}  V^{\text{(3)}}_{ijk},
\end{equation}
where $V^{\text{(2)}}_{ij}$ and $V^{\text{(3)}}_{ijk}$ have three ranges,
\begin{align}
  V^{\text{(2)}}_{ij}
  = & \,
      \sum_{\alpha=1}^3
      V^{\text{(2)}}_\alpha
      \exp\left[- \frac{\left(\vec{r}_i - \vec{r}_j \right)^2}{\mu_\alpha^2} \right]
      \left(W^{\text{(2)}}_\alpha - M^{\text{(2)}}_\alpha P^\sigma P^\tau \right)_{ij},
      \label{2body} \\
  V^{\text{(3)}}_{ijk}
  = & \,
      \sum_{\alpha=1}^3
      V^{\text{(3)}}_\alpha 
      \exp\left[- \frac{\left(\vec r_i - \vec r_j \right)^2}{\mu_\alpha^2}
      -                                                     
      \frac{\left(\vec r_i - \vec r_k\right)^2}{\mu_\alpha^2} \right]
      \notag \\
  & \times 
    \left( W_\alpha^{\text{(3)}} - M_\alpha^{\text{(3)}} P^\sigma P^\tau \right)_{ij} 
    \left( W_\alpha^{\text{(3)}} - M_\alpha^{\text{(3)}} P^\sigma P^\tau \right)_{ik}.
\end{align}
Here, $P^\sigma P^\tau$ represents the exchange of the spin-isospin part of the wave functions
of interacting two nucleons.
The physical coordinate for the $i$th nucleon is $\vec{r}_i$.
The details of the parameters are shown in Ref.~\cite{PhysRevC.49.1814},
but we use F1' parameter set for the Majorana parameter ($M_\alpha^{\text{(3)}}$) of the 
three-body part introduced in Ref.~\cite{PhysRevC.94.064324}.
\par
The G3RS interaction~\cite{PTP.39.91}
is a realistic interaction, and the spin-orbit term has the following form;
\begin{equation}
  \hat{V}_{\text{spin-orbit}}
  =
  \frac{1}{2} \sum_{i \ne j} V^{ls}_{ij}, 
\end{equation}
where
\begin{equation}
  V^{ls}_{ij}
  =
  \left(
    V_{ls}^1 e^{-d_{1} \left( \vec{r}_i - \vec{r}_j \right)^{2}}
    -
    V_{ls}^2 e^{-d_{2} \left( \vec{r}_i - \vec{r}_j \right)^{2}} \right) 
  P\left({}^{3}O\right)
  {\vec{L}}\cdot{\vec{S}}.
  \label{Vls}
\end{equation}
Here, $\vec{L}$ is the angular momentum for the relative motion
between the $i$th and $j$th nucleons, and $\vec{S}$ is the sum of the spin operator
for these two interacting nucleons.
The operator $P\left({}^{3}O\right)$ stands for the projection onto the triplet-odd state.
The strength of the spin-orbit interactions is set to $V_{ls}^1=V_{ls}^2=1800 \, \mathrm{MeV}$,
which allows consistent description of $\nuc{C}{12}$ and $\nuc{O}{16}$~\cite{PhysRevC.94.064324}.
\section{Results}
\label{Results}
\subsection{AQCM basis states}
%
\begin{figure}
  \centering
  \includegraphics[width=5.5cm]{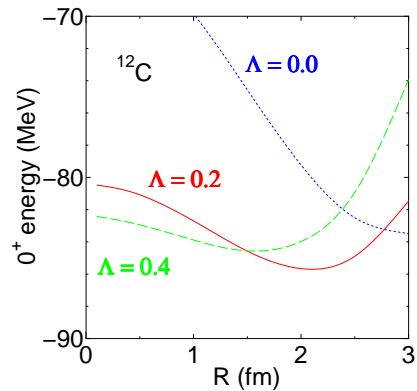} 
  \caption{
    $0^+$ energy curves
    of $\nuc{C}{12}$ calculated with AQCM.
    The horizontal axis
    shows 
    the parameter $R$ representing the distance between the quasi clusters
    with the equilateral triangular configuration.
    The dotted, solid, and dashed curves are the
    cases of $\Lambda$ equal to $0.0$, $0.2$, and $0.4$.}
  \label{aqcm-ec}
\end{figure}
\par
We start the discussion
with the result of AQCM basis states.
Figure~\ref{aqcm-ec} shows
the $0^+$ energy curves
of $\nuc{C}{12}$ calculated with AQCM.
The horizontal axis shows 
the parameter $R$ representing the distance between the quasi clusters
with the equilateral triangular configuration.
The dotted, solid, and dashed curves are the
cases of $\Lambda$ equal to $0.0$, $0.2$, and $0.4$. 
The dotted line ($\Lambda = 0.0$) is the case of three $\alpha$ cluster model
with the equilateral triangular configuration without the 
$\alpha$ breaking effect and resultant spin-orbit contribution,
and the spin-orbit effect is included by setting $\Lambda$ to finite values.
The $\alpha$ cluster model (dotted line)
gives the lowest energy with large $R$ value of $\simeq 3 \, \mathrm{fm}$,
but the spin-orbit interaction strongly lowers AQCM states (finite $\Lambda$) with smaller $R$ values.
However, the finite $\Lambda$ values cause the increase of the kinetic energy,
and the optimal state is obtained as a balance of these two factors. 
The optimal energy of $-86.68 \, \mathrm{MeV}$ is obtained around $R = 2.1 \, \mathrm{fm}$ with $\Lambda=0.2$.
\par
After superposing 18 AQCM basis states
($R = 0.5$, $1.0$, $1.5$, $2.0$, $2.5$, $3.0\,\mathrm{fm}$ and
$\Lambda = 0.0$, $0.2$, $0.4$),
we obtain the lowest $0^+$ state at $-88.04 \, \mathrm{MeV}$,
lower than the energy of the optimal basis state 
($R = 2.1\,\mathrm{fm}$, $\Lambda=0.2$)
by about $1.4\,\mathrm{MeV}$.
\subsection{Inclusion of two-particle-two-hole states}
%
\begin{figure}
  \centering
  \includegraphics[width=5.5cm]{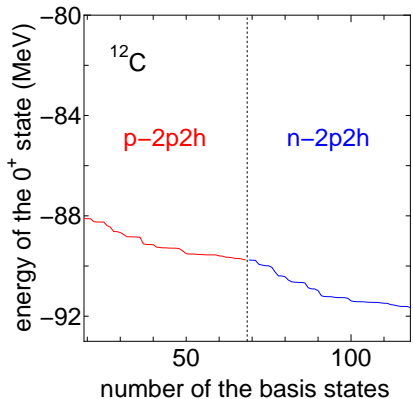} 
  \caption{
    Energy convergence for the $0^+$ state of $\nuc{C}{12}$;
    100 two-particle-two-hole 
    basis states are coupled to the $18$ AQCM basis states.
    The basis states from $19$ to $68$ on the horizontal axis are excited states of the two protons,
    and from $69$ to $118$ are excited states of the two neutrons.}
  \label{c12-118b}
\end{figure}
\par
Then we mix the two-particle-two-hole states to the AQCM basis states.
Figure~\ref{c12-118b} shows the energy convergence for the $0^+$ state of $\nuc{C}{12}$
when we add $100$ two-particle-two-hole
basis states to the $18$ AQCM basis states.
The basis states from $19$ to $68$ on the horizontal axis are excited states of the two protons,
and from $69$ to $118$ are excited states of the two neutrons.
The inclusion of the proton excited states ($69$--$118$ on the horizontal axis)
has an effect of the lowering of the energy by about $2\,\mathrm{MeV}$,
which is quite large.
It is not perfect due to the limitation of the model space, but
the energy is almost converged within the $50$ basis states.
\par
Next, we start superposing the basis states corresponding the two-particle-two-hole
excitation of the neutrons from $69$ on the horizontal axis.
At first, the energy again strongly decreases.
This is because the mixing of the excited states of the neutrons
has the effect of the restoration of the isospin symmetry.
The isospin symmetry is broken when proton excited states are included,
and the broken symmetry is restored by the inclusion of the neutron excited states.
As mentioned in the framework section, the random numbers to
shift of the Gaussian centers of the two nucleons from
the quasi cluster are identical in both cases of proton excitation and neutron excitation.
Thus the model space still contains the room to form 
the isoscalar configuration even after two-particle-two-hole effect is considered;
indeed the isospin symmetry is broken by the Coulomb interaction.
The $0^+$ energy converges to $-91.66\,\mathrm{MeV}$, and the mixing of the two-particle-two-hole
states contributed to the lowering of the ground state energy by more than $3.5\,\mathrm{MeV}$
(the experimental energy of $\nuc{C}{12}$ ground state is $-92.2\,\mathrm{MeV}$).
\subsection{Level spacing of $0^+$ and $2^+$}
%
\begin{figure}
  \centering
  \includegraphics[width=5.5cm]{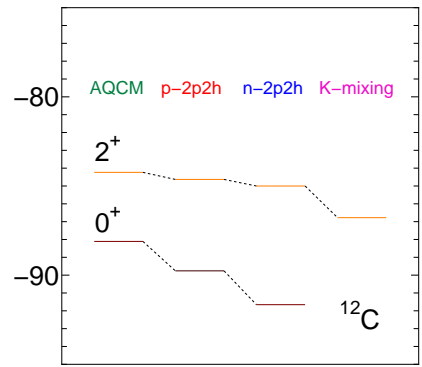} 
  \caption{
    $0^+$--$2^+$ energy spacing of $\nuc{C}{12}$.
    The column ``AQCM'' shows the result obtained after diagonalizing the
    Hamiltonian consisting of the $18$ AQCM  basis states with $K=0$.
    The column ``$p$-2p2h'' shows the result after adding $50$ two-particle-two-hole states for the protons
    within $K=0$.
    In the column ``$n$-2p2h'', $50$ two-particle-two-hole states for the neutrons are mixed,
    where $K$ quantum number is still fixed to $K=0$.
    The mixing of the two-particle-two-hole states allows the $K$-mixing for the $2^+$ state,
    and the effect is shown in the column ``$K$-mixing''.}
  \label{level-space}
\end{figure}
\par
It has been known that traditional $\alpha$ cluster models give
very small level spacing for the ground $0^+$ and first $2^+$ state;
normally the value is about $2$--$3\,\mathrm{MeV}$ compared with the observed value of $4.6\,\mathrm{MeV}$.
It is also known that this defect can be overcome by including the $\alpha$ breaking
effect. The ground state corresponds to the subclosure configuration 
of $p_{3/2}$ in terms of the $jj$-coupling model, and 
the spin-orbit interaction works attractively especially for the $0^+$
state (on the other hand, the excitation to spin-orbit unfavored orbits mixes in the $2^+$ state).
\par
Our result for the $0^+$--$2^+$ energy spacing is summarized in Fig.~\ref{level-space}.
Here, the column ``AQCM'' shows the result obtained after diagonalizing the
Hamiltonian consisting of the 18 AQCM  basis states.
The $0^+$--$2^+$ energy spacing is obtained as $3.9\,\mathrm{MeV}$,
slightly smaller than the experiment.
The AQCM model space only contains the $K=0$ component.
The column ``$p$-2p2h'' shows the result after adding $50$ two-particle-two-hole states for the protons,
where $K$ quantum number is still fixed to $K=0$.
The mixing of $50$ two-particle-two-hole states strongly contributes to the lowering of 
the ground state, and the $0^+$--$2^+$ energy spacing increases to $5.1\,\mathrm{MeV}$, larger than the experiment.
In the column ``$n$-2p2h'', the two-particle-two-hole states for the neutrons are mixed,
where $K$ quantum number is still fixed to $K=0$.
The $0^+$--$2^+$ energy spacing further increases to $6.7\,\mathrm{MeV}$,
quite larger than the experiment.
The result shows that the BCS-like pairing effect is quite important for the $0^+$ state
and increases the level spacing 
between $0^+$ and $2^+$.
However, the mixing of the two-particle-two-hole states allows the $K$-mixing for the $2^+$ state.
The angular momentum projection procedure produces different $K$ states 
($K=1$, $2$) as independent basis states
from each two-particle-two-hole state,
while AQCM basis states ($i=1$--$18$) only contains the $K=0$ component
due to the symmetry of the equilateral triangular ($D_{3h}$) symmetry
even after breaking $\alpha$ clusters. 
After taking into account this $K$-mixing effect,
as shown in the column ``$K$-mixing'', the energy of the $2^+$ state significantly comes down
and finally the spacing becomes 4.9~MeV, quite reasonable value. 
\subsection{Isospin mixing in the ground state}
\par
The $\alpha$ cluster wave function is isoscalar, and this situation is the same
even if we change $\alpha$ clusters to quasi clusters.
However, here we included in the model space the two-particle-two-hole excitation of protons
and neutrons as independent basis states,
thus the isospin symmetry can be broken by the Coulomb interaction
(the nuclear part of the interaction is still isoscalar).
The mixing of the finite isospin can be estimated by the square of the isospin operator.
However, the square of the isospin operator is always constant,
thus here we consider the square of the isospin operator after running
the summation over the particle,
As a result, the operator becomes two-body one,
\begin{equation}
  \hat{O}^{T^2}=\sum_{i,j} \bm{\tau}_i \cdot \bm{\tau}_j,
\end{equation}
where $\bm{\tau}_i$ is the isospin operator for the $i$-th nucleon.
The ground state of the  present model gives the value of $0.016$.
The eigen values of this operator are $0$, $2$, and $6$ for the $T=0$, $T=1$, and $T=2$ state, respectively.
Thus, the present value of $0.016$ means that the isospin is broken at least 
by the order of $10^{-3}$, which is consistent with other calculations.
For instance, the mixing of $T=1$ component in the order of $10^{-4}$ 
in $\nuc{Be}{8}$ is discussed based on the Green's Function Monte Carlo approach~\cite{PhysRevC.88.044333};
however the breaking of the
isospin symmetry is taken into account in the nuclear interaction level there, contrary to the present work.
As mentioned previously, our model space has the room to form the isoscalar configuration,
thus the present result of the isospin mixing is not the numerical artefact. 
\subsection{Principal quantum number}
\par
The physical quantity which reflect the mixing of two-particle-two-hole excitation is required
to confirm the effect. As such candidate,
the expectation value of the principal quantum number $\hat{N}$ of the harmonic oscillator, 
\begin{equation}
  \hat{N} = \sum_i \bm{a}^\dagger_i \cdot \bm{a}_i, 
\end{equation}
can be easily calculated.
Here the summation is over all the nucleons.
The lowest value for $\nuc{C}{12}$ is $8$, corresponding to the state, where four nucleons are in
the lowest $s$ shell and eight nucleons are in the $p$ shell.
The result obtained with the $18$ AQCM basis states give the value of $9.15$,
and after inclusion of the two-particle-two-hole state, the values slightly changes
to $9.13$, but almost identical.
Thus, unfortunately, this quantity cannot be utilized
to discriminate the effect of two-particle-two-hole states.
\subsection{Effect of the proton-neutron pairing}
\par
We have examined the effect of two-particle-two-hole of protons and neutrons.
However, it is known that proton-neutron pairing is quite important in $N=Z$ nuclei~\cite{SATULA19971,
  PhysRevLett.78.3266,
  PhysRevC.87.034310,
  Sagawa_2016}.
We can partially probe this effect; however, it is necessary to reduce the number of 
the basis states for each component of the two-particle-two-hole excitation
from $50$ to $40$ because of the calculation time.
For the basis states corresponding to the proton-neutron pairing,
the Gaussian center parameters of spin-up proton and spin-up neutron 
in one quasi cluster are randomly generated.
%
\begin{figure}
  \centering
  \includegraphics[width=5.5cm]{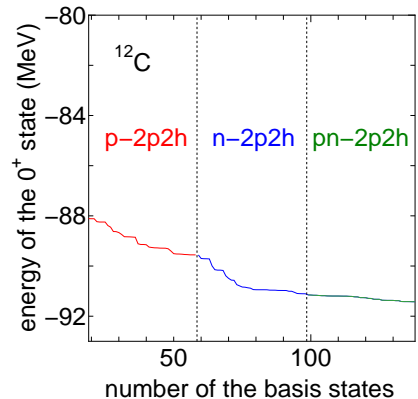} 
  \caption{
    Energy convergence for the $0^+$ state of $\nuc{C}{12}$;
    $120$ two-particle-two-hole 
    basis states are coupled to the $18$ AQCM basis states.
    The basis states from $19$ to $58$ on the horizontal axis are excited states of the two protons,
    from $59$ to $98$ are excited states of the two neutrons, 
    and from $99$ to $118$ are excited states of a proton and a neutron.
  }
  \label{c12-138b}
\end{figure}
\par
The energy convergence for the $0^+$ state of $\nuc{C}{12}$ 
is shown in Fig.~\ref{c12-138b}; $120$ two-particle-two-hole 
basis states are coupled to the $18$ AQCM basis states.
The basis states from $19$ to $58$ on the horizontal axis are excited states of the two protons,
from $59$ to $98$ are excited states of the two neutrons,
and from $99$ to $118$ are excited states of a proton and a neutron.
The number of basis states is not enough and the energy convergence is not perfect;
nevertheless, we can see the basic trend.
Unexpectedly,
the contribution of the proton-neutron excitation is rather limited.
It is considered that the proton-neutron correlations are already included within the
dynamics of the three quasi cluster model.
%
\section{Conclusions} 
\label{Concl}
\par
It has been shown that 
the cluster and single-particle correlations are taken into account in the ground state of $\nuc{C}{12}$.
The recent development of the antisymmetrized quasi cluster model (AQCM) allows us to
generate $jj$-coupling shell model wave functions 
from $\alpha$ clusters models.
The cluster dynamics and the competition with the 
$jj$-coupling shell-model structure can be estimated rather easily.
In the present study, we further included the effect of single-particle excitation;
the mixing of the two-particle-two-hole excited states was considered.
The single-particle excitation
had not always been taken into account in the standard cluster model analyses.
\par
The two-particle-two-hole states are found to strongly contribute to the lowering 
of the ground state owing to the pairing-like correlations.
By extending AQCM, all of the basis states were prepared on the same footing, and they were superposed based on
the framework of GCM.
For the preparation of the two-particle-two-hole states, we used random numbers to
the shift of the Gaussian centers of the two nucleons from
the quasi cluster.
It is stressed that identical sets of random numbers were used in generating the baiss states of both proton excitation and neutron excitation.
Thus, in principle, the model space contains the room to form 
the isoscalar configuration even after two-particle-two-hole effect is considered.
The $0^+$ energy converges to $-91.66\,\mathrm{MeV}$
compared with the experimental value of $-92.2\,\mathrm{MeV}$,
and the mixing of the two-particle-two-hole
states contributed to the lowering of the ground state energy by more than $3.5\,\mathrm{MeV}$.
\par
The isospin symmetry is now broken by the Coulomb interaction,
which can be estimated by the square of the isospin operator.
The ground state of the  present model gives the value of $0.016$.
The eigenvalues of this operator are $0$, $2$, and $6$ for the $T=0$,  $T=1$, and $T=2$ state, respectively.
Thus, the present value of $0.016$ means that the isospin is broken at least 
by the order of $10^{-3}$.
\par
The physical quantity which reflects the mixing of two-particle-two-hole excitation is required
to confirm the effect.
As such a candidate,
the expectation value of the principal quantum number $\hat{N}$ of the harmonic oscillator
was calculated.
The result obtained with the $18$ AQCM basis states gives the value of $9.15$,
and after inclusion of the two-particle-two-hole state, the value slightly changes
to $9.13$, but almost identical. Thus, unfortunately, this quantity cannot be utilized
to discriminate the effect of two-particle-two-hole states.
\par
The proton-neutron pairing is known to play an important in $N=Z$ nuclei,
and
we can prepare proton-neutron two-particle-two-hole states as the basis states, but
unexpectedly, the contribution is rather limited.
It is considered that the proton-neutron correlations are already included within the
dynamics of the three quasi clusters in the present model.
\begin{acknowledgments}
  This work was supported by JSPS KAKENHI Grant Number 19J20543.
  The numerical calculations have been performed using the computer facility of 
  Yukawa Institute for Theoretical Physics,
  Kyoto University. 
\end{acknowledgments}
%
%
%
\end{document}